\documentclass[aps,prd,a4paper,twocolumn,amsmath,showpacs,superscriptaddress,nofootinbib,preprintnumbers]{revtex4-1}

\usepackage{verbatim}
\usepackage[T1]{fontenc}
\usepackage[utf8]{inputenc}
\usepackage[american]{babel}
\usepackage{epsfig}
\usepackage{graphicx}
\usepackage{booktabs}
\usepackage{multirow}
\usepackage{dcolumn}
\usepackage{amsmath}
\usepackage{mathtools}
\usepackage{amsfonts}
\usepackage{amssymb}
\usepackage{epstopdf}
\usepackage{bm}
\usepackage{siunitx}
\usepackage{braket}
\usepackage{enumitem}
\usepackage{soul}
\usepackage[table]{xcolor}
\usepackage{color}
\usepackage{transparent}
\usepackage{pifont}

\definecolor{navyblue}{rgb}{0.0, 0.0, 0.5}
\definecolor{royalblue}{rgb}{0.25, 0.41, 0.88}
\definecolor{cadmiumgreen}{rgb}{0.0, 0.42, 0.24}
\definecolor{blue-violet}{rgb}{0.54, 0.17, 0.89}
\definecolor{darkviolet}{rgb}{0.58, 0.0, 0.83}
\definecolor{orange(colorwheel)}{rgb}{1.0, 0.5, 0.0}

\usepackage{hyperref}
\hypersetup{
    colorlinks=true, 
    linkcolor=royalblue, 
    citecolor=magenta}

\newcommand\ee{\end{equation}}
\newcommand\be{\begin{equation}}
\newcommand\eea{\end{eqnarray}}
\newcommand\bea{\begin{eqnarray}}

\newcommand{\neff}{N_{\rm eff}}

\renewcommand{\vec}{\bm}

\usepackage{booktabs}
\usepackage{multirow}
\usepackage{dcolumn}
\usepackage{colortbl}

\newcommand\horsp{\rule[-1.5mm]{0mm}{4.125mm}}
\newcommand\morehorsp{\rule[-2.25mm]{0mm}{6mm}}

\definecolor{magenta(process)}{rgb}{1.0, 0.0, 0.56}

\definecolor{darkspringgreen}{rgb}{0.09, 0.45, 0.27}

\definecolor{royalblue(web)}{rgb}{0.25, 0.41, 0.88}

\newcommand{\vtheta}{{\boldsymbol{\theta}}}

\newcommand{\vx}{\vec{x}}

\begin{document}

\title{Bayesian Evidence against Harrison-Zel'dovich spectrum in tension cosmology}  

\author{Eleonora Di Valentino}
\email{eleonora.divalentino@manchester.ac.uk}
\affiliation{Jodrell Bank Center for Astrophysics, School of Physics and Astronomy, University of Manchester, Oxford Road, Manchester, M13 9PL, UK}
\author{Alessandro Melchiorri}
\email{alessandro.melchiorri@roma1.infn.it}
\affiliation{Physics Department and INFN, Universit\`a di Roma ``La Sapienza'', Ple Aldo Moro 2, 00185, Rome, Italy} 
\author{Yabebal Fantaye}
\email{yabi@aims.ac.za}
\affiliation{African Institute for Mathematical Sciences,6-8 Melrose Road, Muizenberg 7945, South Africa}
\affiliation{Department of Mathematics, University of Stellenbosch, Stellenbosch 7602, South Africa} 
\author{Alan Heavens}
\email{a.heavens@imperial.ac.uk}
\affiliation{Imperial Centre for Inference and Cosmology (ICIC),
Imperial College, Blackett Laboratory,
Prince Consort Road, London SW7 2AZ, U.K.}

\date{\today}
\begin{abstract}
Current cosmological constraints on the scalar spectral index of primordial fluctuations $n_{\rm s}$ in the $\Lambda$CDM model have excluded the minimal scale-invariant Harrison-Zel'dovich model ($n_{\rm s}=1$; hereafter HZ) at high significance, providing support for inflation. In recent years, however, some tensions have emerged between different cosmological datasets that, if not due to systematics, could indicate the presence of new physics beyond the $\Lambda$CDM model. In the light of these developments, we evaluate the Bayesian evidence against HZ in different data combinations and model extensions. Considering only the Planck temperature data, we find inconclusive evidence against HZ when including variations in the neutrino number $N_{\rm eff}$ and/or the Helium abundance $Y_{\rm He}$. Adding the Planck polarization data, on the other hand, yields strong evidence against HZ in the extensions we considered.  Perhaps most interestingly, 
Planck temperature data combined with local measurements of the Hubble parameter \citep{R16,riess2018} 
give as the most probable model 
an HZ spectrum, with additional neutrinos.  However, with the inclusion of polarisation, standard $\Lambda$CDM is once again preferred, but the HZ model with extra neutrinos is not strongly disfavored.  The possibility of fully ruling out the HZ spectrum is therefore ultimately connected with the solution to current tensions between cosmological datasets. If these tensions are confirmed by future data, then new physical mechanisms could be at work and an HZ spectrum could still offer a valid alternative.
\end{abstract}

\maketitle

\section{Introduction}

Current observations of Cosmic Microwave Background (CMB) anisotropies and Large Scale Structure are in good agreement with the hypothesis that cosmic structures originated from tiny density perturbations in the early universe.
The inflationary theory (see e.g. \cite{reviews} for reviews) predicts the existence of such perturbations by stretching microscopic quantum fluctuations to cosmological scales \cite{perturbations}. While the exact inflationary mechanism by which these perturbations are generated is not yet known, a general prediction is that their power spectrum can be well described by a power law $A_{\rm s} k^{n_{\rm s}}$ where $A_{\rm s}$ and $n_{\rm s}$ are defined as the primordial amplitude and spectral index while $k$ is the perturbation wavenumber measured in Mpc$^{-1}h$. Furthermore, the value of the spectral index should be nearly one, $n_{\rm s}\sim1$, reflecting the constancy of the Hubble horizon during inflation, but at the same time not exactly one, due to the dynamics of the inflaton field (again, see \cite{perturbations}).

An exact value of $n_{\rm s} = 1$ is indeed not expected in inflation and would coincide with the phenomenological model proposed by Harrison \cite{HZ1}, Zel’dovich \cite{HZ2}, and Peebles and Yu \cite{HZ3}, known as Harrison-Zel'dovich (HZ) spectrum, proposed well before the formulation of inflation, and corresponding to perfect scale-invariance of the fluctuations. 
While it is still possible to have inflationary models with spectral index nearly identical to HZ (see e.g. \cite{fakehz}), a measurement of $n_{\rm s}$ close but different from one should be considered as a further corroboration of inflation.

In the past twenty years, CMB measurements made by balloon experiments such as BOOMERanG \cite{deBernardis:2000sbo1,deBernardis:2000sbo2} and satellite experiments such as WMAP \cite{wmap1,wmap2} and, more recently, Planck \cite{planck2013,planck2015}, have provided improving constraints on $n_{\rm s}$. From the constraint of $n_{\rm s}=0.90\pm0.08$ at $68 \%$ credible interval from BOOMERanG \cite{deBernardis:2000sbo2}, we have now $n_{\rm s}=0.9645\pm0.0049$  from the Planck 2015 data release, i.e. an increase by a large factor of $\sim 16$ in the precision of the measurement and a preference over the HZ spectrum at about $7$ standard deviations.

This is a success for the theory of inflation and several CMB experiments are now aiming at the measurement of polarization $B$ modes generated by gravitational waves during inflation (see e.g. \cite{kamionkowski}). 

It is important to stress, however, that the above constraints have been obtained indirectly, assuming the $\Lambda$CDM model based on Cold Dark Matter (CDM) and a cosmological constant ($\Lambda$).
Moreover, the unprecedented sensitivity in cosmological experiments is revealing several interesting discrepancies and tensions in the $\Lambda$CDM model. 

For example, the Planck constraint on the Hubble constant, obtained under $\Lambda$CDM, is about $3.3$ standard deviation from the direct constraint of Riess et al. 2016 \cite{R16} (R16 hereafter), derived from direct observations. The disagreement is even larger, $3.8$ standard deviations, for the new determination of Riess et al. 2018 \cite{riess2018}. Furthermore, the Planck temperature anisotropy power spectrum data seems to suggest an amplitude of gravitational lensing larger than the one expected in the $\Lambda$CDM scenario at about $\sim 2-2.5$ standard deviations (\cite{planck2015,plancklike,planckshift,hulensing}), showing a possible internal tension in the Planck data itself. A greater amount of lensing in the Planck power spectra, parametrized by the $A_{\rm lens}$ factor (see \cite{calens}), puts the Planck cosmology in better agreement with the cosmic shear data from surveys such as the Kilo-degree survey KiDS-450 \cite{kids} and the Dark Energy Survey (DES) \cite{deswl1,deswl2}, as well as with the cosmological parameters derived from WMAP data \cite{bennett}. 

While the statistical significance of these tensions is mild \cite{heavens}, the possibility of extensions to the $\Lambda$CDM scenario that could explain them is clearly open. For example, an increase in the number density of relativistic particles at recombination $N_{\rm eff}$ or a change in the dark energy equation of state $w$ could both alleviate the current discrepancy on the Hubble parameter (see e.g. \cite{divalentino}). In the past years the possibility of new physics either in the dark energy sector either in the neutrino sector to solve the Hubble tension has been considered in several works (\cite{divalentino,Bernal:2016gxb,Zhao:2017cud,Archidiacono:2016kkh,
Ko:2016uft,DiValentino:2017iww,Qing-Guo:2016ykt,Chacko:2016kgg,
Lin:2017bhs,Sola:2017znb,Karwal:2016vyq,Brust:2017nmv,Prilepina:2016rlq,
Yang:2017amu,Zhao:2017urm,Zhang:2017idq,DiValentino:2017rcr,DiValentino:2017oaw}).

It is therefore timely to investigate the robustness of the conclusion that the HZ spectrum is ruled out while considering extended cosmological scenarios, beyond $\Lambda$CDM. A similar analysis has been already performed in recent papers (see e.g. \cite{Benetti:2017juy,Benetti:2017gvm,Benetti:2013wla,Pandolfi:2010dz}). Here we extend these previous analyses by including more data (for example, the Planck polarization CMB data), by considering more parameter extensions and by using a different approach in calculating Bayesian evidence using the MCEvidence code described in \cite{fantaye}. Moreover, when computing Bayesian evidence we will compare the viability of the HZ spectrum not only with respect to  $\Lambda$CDM but also to its extensions.

As we will see, a crucial point in this investigation is that an HZ model has $n_{\rm s}=1$, i.e. one parameter fewer than standard $\Lambda$CDM. The HZ model is therefore less complicated (from the point of view of the number of parameters) and this may lead to a higher Bayesian Evidence when compared with models where $n_{\rm s}$ is an additional parameter and which produce similar fits to the data. Indeed, Bayesian Evidence weights the simplicity of the model with the Occam factor, the inverse factor by which the prior space collapses when the data arrive.

The paper is structured as follows: in the next section we describe the data analysis method, in Section III we discuss the results and in Section IV we present conclusions.

\section{Method}

\subsection{Models considered}

As stated in the introduction the goal of this paper is to determine the Bayesian Evidence for an HZ spectrum in $\Lambda$CDM and extended scenarios.
We have therefore analyzed the cosmological data under the assumption of the following models:

\begin{itemize}

\item Standard $\Lambda$CDM. In this case we assume a flat model, with cold dark matter, a cosmological constant and adiabatic primordial fluctuations. For this model we have considered variations in $6$ parameters: the amplitude $A_{\rm s}$ and spectral index $n_{\rm s}$ of primordial scalar fluctuations, the cold $\omega_{\rm c}$ and baryonic $\omega_{\rm b}$ matter densities, the angular size of the acoustic horizon at decoupling $\theta_{\rm c}$ and the reionization optical depth $\tau$.

\item $\Lambda$CDM+$N_{\rm eff}$. In this case we have extended the $\Lambda$CDM model described above by including variation in the neutrino effective number $N_{\rm eff}$ that essentially counts the number of relativistic degrees of freedom at recombination. Standard model with three  neutrinos of negligible mass predicts $N_{\rm eff}=3.046$. 
We assume a flat prior on $N_{\rm eff}$ between $0.05$ and $10$. The inclusion of $N_{\rm eff}$ affects the CMB constraints on $n_{\rm s}$ (see e.g. \cite{Benetti:2013wla}).

\item $\Lambda$CDM+$Y_{\rm He}$. Varying the Helium abundance $Y_{\rm He}$ modifies the process of recombination and changes the structure of peaks in the CMB anisotropy spectra. This quantity is usually derived from the value of the baryon density $\omega_{\rm b}$ assuming standard Big Bang Nucleosynthesis (BBN). However it is plausible to take a more model-independent approach and to derive constraints on $Y_{\rm He}$ directly from CMB observations. The assumed prior on $Y_{\rm He}$ is flat between $0.1$ and $0.5$.

\item $\Lambda$CDM+$N_{\rm eff}$+$Y_{\rm He}$. In this case we remove completely the assumption of BBN and of the standard three neutrino framework and consider both the possibilities of an extra background of relativistic particles and free $Y_{\rm He}$.

\item $\Lambda$CDM+$N_{\rm eff}$+$n_{\rm run}$+$\Sigma m_{\nu}$+$A_{\rm lens}$. The model described above is further extended by considering the possibility of a running of the spectral index with scale $n_{\rm run}=dn_{\rm s}/d\ln k$, a total neutrino mass $\Sigma m_{\nu}$, and a varying amplitude of the CMB lensing signal $A_{\rm lens}$. In what follows we will refer to this model as {\bf Extended-$10$} since we consider $10$ free parameters.  The prior on $n_{\rm run}$ is flat between $-1$ and $1$. The prior on $m_{\nu}$ is flat between $0$ and $5$eV. The prior on $A_{\rm lens}$ is flat between $0$ and $10$.

\item $\Lambda$CDM+$N_{\rm eff}$+$n_{\rm run}$+$\Sigma m_{\nu}$+$A_{\rm lens}$+$w$. We further extend the Extended-$10$ model by considering variations in the dark energy equation of state $w=p/(\rho c^2)$, assumed to be constant with redshift. We will refer to this model as {\bf Extended-$11$}.
The prior on $w$ is flat between $-3$ and $0.3$.

\end{itemize}

The inclusion of $N_{\rm eff}$ and $w$ is motivated by a well known parameter degeneracy with the value of the Hubble constant derived from the Planck data. Increasing $N_{\rm eff}$ or decreasing $w$ could bring the Planck constraint on $H_0$ in better agreement with the direct measurement of $H_0$ from R16 \cite{R16}. We consider variation in $A_{\rm lens}$ given the indication from the Planck data for an anomalous $A_{\rm lens}>1$ value. We also include $n_{\rm run}$ and $Y_{\rm He}$ since these parameters are correlated with $n_{\rm s}$.

A few remarks about other parameters in is order.  Aside from cosmological parameters, the Planck analysis also includes a number of nuisance parameters.  These are marginalised over before the evidence is computed, which is a valid procedure if the nuisance parameters are independent of the cosmological parameters. It is a good approximation for Planck \citep{planck2015XI}.

Finally, we note that, for uniform priors, the Bayesian Evidence depends inversely on the prior range, provided that the prior encompasses all of the likelihood.  This makes it very straightforward to recalculate the Bayes factors for different prior ranges, if desired.

\begin{figure*}
\includegraphics[width=1\textwidth]{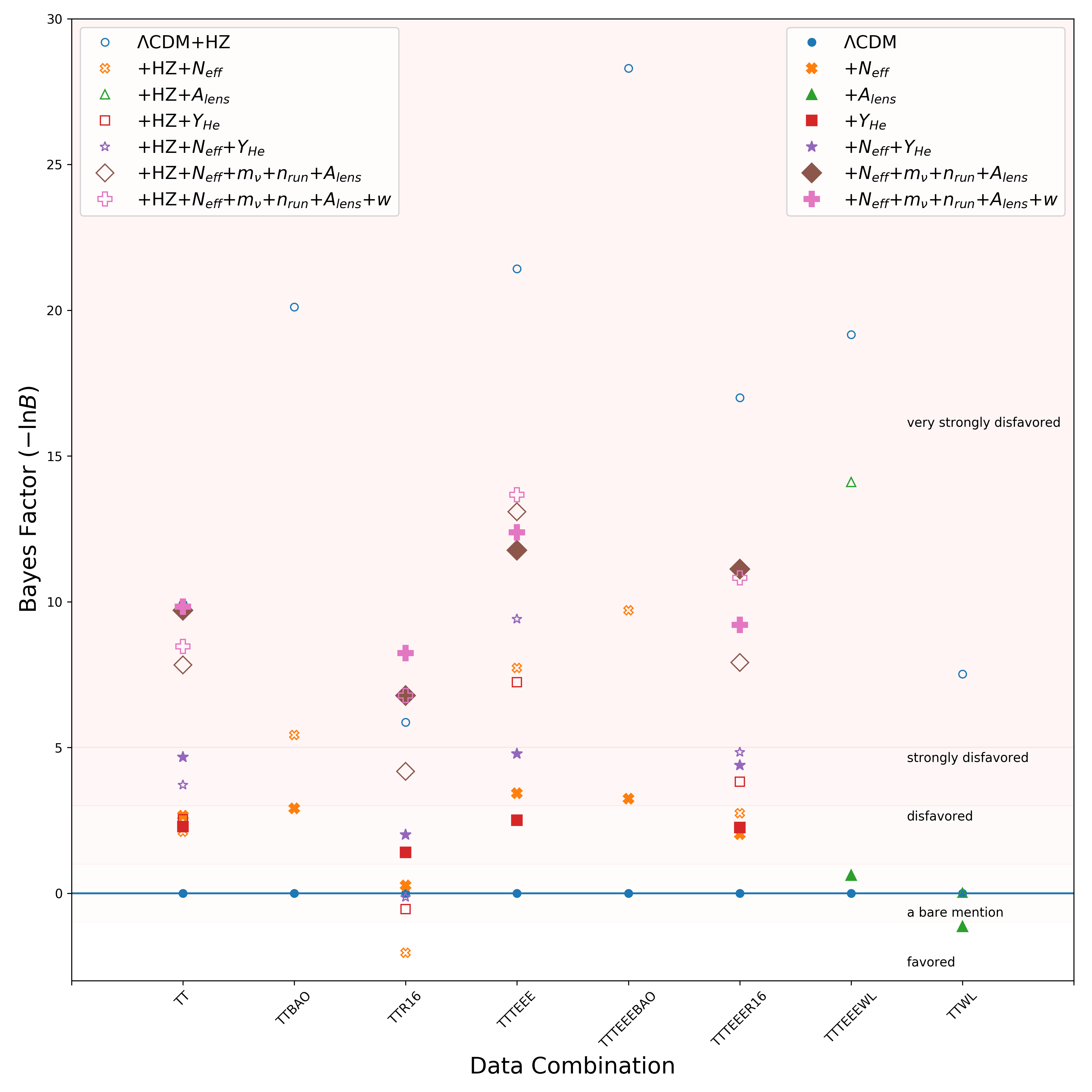}
\caption{Bayes factors $ - \ln B$ w.r.t. the flat $\Lambda$CDM model. Following our definition, a negative value (please note the minus sign on the $y$ label) provides evidence against an HZ spectrum while a positive value favors it. Models with varying spectral index ($n_{\rm s} \neq 1$) are shown by filled markers while the Harrison-Zel'dovich ($n_{\rm s}=1$) cases are shown by the open markers. The different models shown in the bottom left legend are extensions of the flat $\Lambda$CDM model while those in the bottom right legend are extensions of the flat $\Lambda$CDM with HZ spectral index. The number of parameters in the model is represented by the relative size of the markers. The colored boundaries delineating the evidence degrees are based on the Kass \& Raftery (1995) scale. Note that in case of the BAO and WL datasets we consider just $N_{\rm eff}$ and $A_{\rm lens}$ respectively as extra parameters.
}\label{fig}
\end{figure*}

\subsection{Data}

As cosmological data we examine the high-$\ell$ temperature and low-$\ell$ temperature and polarization CMB angular power spectra released by Planck in 2015 \cite{plancklike}. We consider different set of data combinations. The first set includes the large angular-scale temperature and polarization anisotropies measured by the Planck LFI experiment and the small-scale temperature anisotropies measured by Planck HFI experiment, we refer this case by ``Planck TT''.
The second set includes Planck TT together with the high-$\ell$ polarization data measured by Planck HFI \cite{plancklike}, and this dataset is refereed  as ``Planck TTTEEE''. 
We also include the R16 bound in the form of an additional Gaussian likelihood weighting for the Hubble constant $H_0=73.24\pm1.75$ km s$^{-1}$ Mpc$^{-1}$ at $68 \%$ credibility interval, as measured by \cite{R16}.

Finally, in some cases we will also use information from Baryon Acoustic Oscillation (BAO) and cosmic shear weak lensing (WL) surveys as in \cite{planck2015}. 

The data are first analyzed using the November 2016 version of the publicly available Monte Carlo Markov Chain package \texttt{cosmomc} \cite{Lewis:2002ah} with a convergence diagnostic based on the Gelman and Rubin statistic (see \url{http://cosmologist.info/cosmomc/}).  The MCMC chains in the Planck legacy archive are described at \verb!https://wiki.cosmos.esa.int/planck-legacy-!\\
\verb!archive/index.php/Cosmological_Parameters!.

\subsection{Bayesian Evidence}

In this paper we compare models principally using the framework of Bayesian Evidence. The posterior probability of a model $M$ given the data $\vx$, $p(M|\vx)$ depends on the Bayesian Evidence (or marginal likelihood), $p(\vx| M)$,which is the denominator in the posterior for a vector of parameters $\vtheta$ of a model $M$ and a set of data $\vx$:
\be
p(\vtheta|\vx, M) = \frac{p(\vx|\vtheta, M)\,\pi(\vtheta|M)}{p(\vx|M)}.
\ee
Here $p\left(\vx | \vtheta, M\right)$ is the likelihood and $\pi\left(\vtheta | M\right)$ is an assumed prior on the parameters.

The Bayesian Evidence ensures that the posterior is normalised, and is given by
\be
E \equiv p(\vx|M) = \int d\vtheta\, p(\vx|\vtheta,M)\,\pi(\vtheta|M).
\label{eq: Marginal Equation}
\ee

In the light of data $\vx$, Bayesian model comparison proceeds by pairwise comparison of competing models, say $M_0$ and $M_1$, through their posterior odds ratio:
\be
\frac{p(M_0 | \vx)}{p(M_1 | \vx)} =  \frac{p(\vx|M_0)}{p(\vx|M_1)}\,\frac{\pi(M_0)}{\pi(M_1)}.
\ee
Assuming equal prior probabilities for the competing models, $\pi(M_0)=\pi(M_1)$, the models' posterior odd ratio is the Bayes factor, 
\be 
{\cal B} \equiv \frac{p(\vx|M_0)}{p(\vx|M_1)}.
\ee 

According to the revised Jeffreys scale by Kass and Raftery \cite{kassraftery}, the evidence (against $M_1$) is considered as {\em positive} if $1.0 < \ln {\cal B} < 3.0$, {\em strong} if $3.0 < \ln {\cal B} <5.0$, and {\em very strong} if  $\ln {\cal B} > 5.0$.

In what follows we will always consider the evidence {\it against} an HZ model, i.e. $M_0$ is a model with varying $n_{\rm s}$ parameter, while $M_1$ is a model with an HZ spectrum. Following this definition a positive value of $\ln {\cal B}$ provides evidence {\it against} an HZ spectrum. A negative value of $\ln {\cal B}$ means evidence against $n_{\rm s} \neq 1$ model. 

The evidence is computed from the MCMC chains using the MCEvidence code described in \cite{fantaye}.

\section{Results}

\begin{table*}
\begin{center}\footnotesize
\begin{tabular}{l|c|c|c|c}
\toprule
\horsp
Model & Planck TT&Planck TT$+R16$&Planck TTTEEE&Planck TTTEEE$+R16$\\
\hline
$\Lambda$CDM (HZ)&$9.94$&$5.86$&$21.42$&$16.99$ \\
\morehorsp
$\Lambda$CDM+$N_{\rm eff}$(HZ)&$2.11$&$-2.04$&$7.73$&
$2.75$\\
\morehorsp
$\Lambda$CDM+$Y_{\rm He}$ (HZ)& ${2.53}$&$-0.55$&$7.24$ &$3.82$ \\
\morehorsp
$\Lambda$CDM+$Y_{\rm He}$+$N_{\rm eff}$ (HZ)& $3.71$ &$-0.14$&$9.4$&$4.83$ \\
\morehorsp
Extended-$10$ (HZ) & $7.85$& $4.18$&$13.08$&
$7.91$\\ 
\morehorsp
Extended-$11$ (HZ)& $8.48$&$6.78$&$13.67$&$10.82$ \\ 
\bottomrule
\end{tabular}
\end{center}
\caption{Bayesian Evidences against an HZ spectrum under different model assumptions with respect to the standard $\Lambda$CDM model with $n_{\rm s}$ free to vary.}
\label{Tablenew1}
\end{table*}

\begin{table*}
\begin{center}\footnotesize
\begin{tabular}{l|c|c|c|c}
\toprule
\horsp
Model & Planck TT&Planck TT$+R16$&Planck TTTEEE&Planck TTTEEE$+R16$\\
\hline
\morehorsp
$\Lambda$CDM+$N_{\rm eff}$(HZ)&$-0.55$&$-2.31$&$4.30$&
$0.72$\\
\morehorsp
$\Lambda$CDM+$Y_{\rm He}$ (HZ)& $0.25$&$-1.95$&$4.74$ &$1.57$ \\
\morehorsp
$\Lambda$CDM+$Y_{\rm He}$+$N_{\rm eff}$ (HZ)& $-0.95$ &$-2.04$&$3.79$&$0.44$ \\
\morehorsp
Extended-$10$ (HZ) & $-1.86$& $-2.60$&$1.32$&
$-3.2$\\ 
\morehorsp
Extended-$11$ (HZ)& $-1.35$&$-1.45$&$1.29$&$1.61$ \\ 
\bottomrule
\end{tabular}
\end{center}
\caption{Bayesian Evidences against an HZ spectrum under different model assumptions but now comparing with the corresponding model extension with $n_s$ free. For example, the HZ spectrum under $\Lambda$CDM$+N_{\rm eff}$ in the first line is compared with the corresponding $\Lambda$CDM$+N_{\rm eff}$ model but with $n_{\rm s}$ free to vary. }
\label{Tablenew2}
\end{table*}

Before discussing in detail all the obtained results in the next section it is useful to consider Figure~\ref{fig} where we plot the Bayes factors considering several data combinations and different theoretical scenarios. The Bayes factors for each dataset are with reference to the $\Lambda$CDM case, solid symbols identify a model where $n_{\rm s}$ is allowed to vary while empty symbols correspond to models where an HZ spectrum is assumed. 
If we first consider models with free $n_{\rm s}$ (solid symbols) we notice that there is no parameter extension that is favored with respect to $\Lambda$CDM with the only exception of the $\Lambda$CDM+$A_{\rm lens}$ model (solid green triangle) with just a minor, positive, evidence for the Planck+WL dataset. This is a direct consequence of the anomalous $A_{\rm lens}$ value seen by the Planck data.
We can also notice a strong and a very strong evidence against Extended-$10$ and Extended-$11$ (solid brown diamond and solid pink cross) with respect to $\Lambda$CDM, especially in the case of the Planck TTTEEE dataset. Models with one single additional parameter as $\Lambda$CDM+$N_{\rm eff}$ (red solid squares) or $\Lambda$CDM+$Y_{\rm He}$ (orange solid cross) are not strongly disfavored. In practice the visual fact that most of the models are below the blue line clearly indicates that there is currently no strong evidence against the $\Lambda$CDM standard scenario.

When moving to empty symbols, i.e. to models that now assume an HZ spectrum, we also see that there is no positive evidence for them with the single notable exception of the Planck TT+R16 dataset. Indeed, in this case we see a positive evidence with respect to $\Lambda$CDM for HZ $\Lambda$CDM+$N_{\rm eff}$ (empty orange times) and HZ $\Lambda$CDM+$Y_{\rm He}$ (empty red square). The positive evidence for the HZ $\Lambda$CDM+$A_{\rm lens}$ is still marginally present for the Planck TT+WL dataset but disappears completely for Planck TTTEEE +WL case, with a very strong negative evidence. We also notice strong or very strong evidence against HZ in the $\Lambda$CDM model for all datasets considered. 
Very strong evidence against HZ with respect to $\Lambda$CDM is also present for all model extensions considered in the case of the Planck TTTEEE datasets. 
Visually we see that the Planck TT + R16 dataset provides the least stringent constraints on HZ and that the inclusion of R16 reduces the evidence against HZ for Planck TTTEEE.

In the next sections we discuss these results in more detail provide the Bayesian evidences for several data and model combinations.  The constraints on cosmological parameters  can be found in the Appendix to this paper. 

\subsection{Planck data and the R16 constraint}

In Table~\ref{Tablenew1} we compute the evidence for an HZ spectrum for several model extensions with respect to the standard $\Lambda$CDM model, i.e. the quantity

\be 
\ln {\cal B} \equiv \ln \frac{p(\vx|M_{\Lambda CDM,n_s\neq1})}{p(\vx|M_{1,n_s=1})}.
\ee 

\noindent where $M_{\Lambda CDM}$ is standard $\Lambda$CDM with variable $n_{\rm s}$ and $M_1$ is one of the models listed in the first column of Table~\ref{Tablenew1}
with an HZ primordial spectrum, i.e. $n_{\rm s}=1$. The evidences are computed assuming the Planck CMB data with and without the inclusion of the $R16$ constraint. 

We can immediately notice  (first row of Table~\ref{Tablenew1}) that an HZ spectrum is strongly disfavored with a very strong negative evidence ($\ln {\cal B}>5$) in case of $\Lambda$CDM for any data combination. In the framework of $\Lambda$CDM, an HZ spectrum is therefore significantly ruled out. This is clearly in agreement with the accurate constraint that the Planck data provides on the scalar spectral index when a $\Lambda$CDM model is assumed (see the results in Tables ~\ref{tab:TT},~\ref{tab:TT+R16},
~\ref{tab:TTTEEE}, and ~\ref{tab:TTTEEE+R16} in the Appendix).

However, when we consider extensions involving $N_{\rm eff}$ or $Y_{\rm He}$ but with an HZ spectrum (rows $2-4$ of Table~\ref{Tablenew1}), the Planck TT data alone do not significantly prefer standard $\Lambda$CDM over these models providing just a negative evidence ($\ln {\cal B}>2$). Furthermore, when the $R16$ constraint is included with TT, model extensions with an HZ spectrum are even favored with respect to standard $\Lambda$CDM with positive, albeit not significantly large, evidences ($\ln {\cal B} <0$).
In practice both the $\Lambda$CDM$+N_{\rm eff}$ and $\Lambda$CDM$+Y_{\rm He}$ models with $n_s=1$, provide a better fit to the Planck TT $+R16$ dataset than standard $\Lambda$CDM with the same number of parameters ($6$) (see discussion on this point in the Appendix). 

The inclusion of CMB polarization data, however, lifts some of the parameter degeneracies that affect the CMB temperature data, provides a better constraint on $N_{\rm eff}$ and $Y_{\rm He}$ compatible with the expected standard values, and disfavors an HZ spectrum in these model extensions.  Indeed, very strong evidence ($\ln {\cal B}>5$) against an HZ spectrum for all the model extensions considered with respect to standard $\Lambda$CDM is obtained with the Planck TTTEEE data. Once the R16 data are included, the evidence is still present against HZ in model extensions that varies $N_{\rm eff}$ or $Y_{\rm He}$ but only at the level of $\ln {\cal B}>2$. An HZ spectrum in these models is therefore disfavored but not fully excluded with respect to $\Lambda$CDM when the Planck TTTEEE + R16 dataset is considered. For the same dataset, the very extended models Extended$-10$ or Extended$-11$ with an HZ spectrum are strongly disfavored with respect to $\Lambda$CDM ($\ln {\cal B}>5$).

It is interesting to compute the evidence for an HZ spectrum not with respect to standard $\Lambda$CDM but considering the same model but with $n_s$ free to vary, i.e.:

\be 
\ln {\cal B} \equiv \ln \frac{p(\vx|M_{1,n_s\neq1})}{p(\vx|M_{1,n_s=1})}.
\ee 

\noindent where $M_1$ is one of the of the extensions to $\Lambda$CDM.

The results of this kind of analysis are reported in Table~\ref{Tablenew2}. As we can see, when considering extensions to $\Lambda$CDM, there is no very strong evidence against an HZ spectrum (all the values in the Table are $<5$). In particular, we always found a positive evidence for HZ in the Planck TT+R16 dataset and a marginally negative or positive evidence in the case for Planck TTTEEE+R16. In short, when considering model extensions an HZ spectrum is never significantly ruled out from CMB data alone and is in some cases even favored when the R16 constraint is included.
Therefore, if the current case for extensions motivated by the tensions between the Planck and the R16 results on the Hubble constant will be further confirmed by future data the HZ spectrum could be still a viable option for the primordial density perturbations.

\begin{table*}
\begin{center}\footnotesize
\begin{tabular}{|l|c|c|}
\hline
Planck TT+BAO& $\Lambda$CDM&$\Lambda$CDM+$N_{\rm eff}$\\
\hline
$\Lambda$CDM (HZ)& $20.12$ & $17.08$\\ 
$\Lambda$CDM+$N_{\rm eff}$ (HZ)& $5.43$ & $2.52$\\ 
\hline
\hline
Planck TTTEEE+BAO&$\Lambda$CDM&$\Lambda$CDM+$N_{\rm eff}$\\
\hline\morehorsp
$\Lambda$CDM (HZ)& $28.3$ & $25.05$\\ 
$\Lambda$CDM+$N_{\rm eff}$ (HZ)& $9.71$ & $6.46$\\ 
\hline
\end{tabular}
\end{center}
\caption{Bayesian evidence for an HZ spectrum in case of $\Lambda$CDM and $\Lambda$CDM+$N_{\rm eff}$. Planck TT+BAO and Planck TTTEEE+BAO data are considered.}
\label{tablenew3}
\end{table*}

\subsection{Planck + BAO}

In the previous section we have considered the combination of Planck data with the R16 constraint. The R16 constraint on the Hubble constant is in tension with the corresponding Planck constraint obtained standard $\Lambda$CDM. We have therefore seen that if we assume this tension as genuine and not produced by unknown systematics in the data then there is no significant evidence ($\ln {\cal B}>5$) against an extended model with an HZ spectrum either with respect to $\Lambda$CDM, or to the extension itself with $n_s\neq1$.

However, other datasets such as BAO are in better agreement with Planck when $\Lambda$CDM is assumed and it is interesting to evaluate the evidence against HZ when these two datasets are combined. 

In Table~\ref{tablenew3} we report the Bayesian evidence for HZ for Planck TT+BAO and Planck TTTEEE+BAO data, considering  for simplicity just the $\Lambda$CDM$+N_{\rm eff}$ extension. Indeed, this extension seems to provide the best solution to the $H_0$ tension. In the second column of Table~\ref{tablenew3} we provide the evidence against the model with HZ listed in the first column with respect to standard $\Lambda$CDM. In the third column of Table~\ref{tablenew3} we report the similar evidence but now with respect to $\Lambda$CDM$+N_{\rm eff}$ with $n_s$ free to vary.

As we can see (second column), Planck+BAO always provides strong evidence against HZ with respect to standard $\Lambda$CDM. When the BAO data are included, the evidence against HZ under $\Lambda$CDM grows by $\Delta \ln {\cal B} = 10.18$ for Planck TT and by $\Delta \ln {\cal B} = 6.9$ for Planck TTTEEE. When considering an HZ spectrum in a $\Lambda$CDM$+N_{\rm eff}$ extension, the evidence against it with respect to standard $\Lambda$CDM also grows by $\Delta \ln {\cal B} = 3.33$ for Planck TT and by $\Delta \ln {\cal B} = 1.98$ for Planck TTTEEE. While HZ was already ruled out from Planck TTTEEE data alone, the inclusion of the BAO datasets excludes HZ also in the case of Planck TT.

When we consider the evidence with respect to $\Lambda$CDM$+N_{\rm eff}$
(third column) we first note that $\Lambda$CDM models with an HZ spectrum are significantly disfavored ($\ln {\cal B} >5$) both from Planck TT+BAO and Planck TTTEEE+BAO, even being based on fewer free parameters ($5$ instead of $7$). Interestingly, there is no significant evidence against an HZ spectrum when the $\Lambda$CDM$+N_{\rm eff}$ model is considered from Planck TT+BAO data even if there is
an increase of $\Delta \ln {\cal B} = 3.07$ with respect to the Planck TT case.
Finally, we see that inclusion of the BAO data with Planck TTTEEE data provides now a very strong evidence against HZ also in the $\Lambda$CDM$+N_{\rm eff}$ scenario.

\begin{table*}
\begin{center}\footnotesize
\begin{tabular}{|l|c|c|}
\hline
Planck TT+WL& $\Lambda$CDM&$\Lambda$CDM+$A_{\rm lens}$\\
\hline
$\Lambda$CDM (HZ)& $7.51$ & $8.65$\\ 
$\Lambda$CDM+$A_{\rm lens}$ (HZ)& $0.01$ & $1.15$\\ 
\hline
\hline
Planck TTTEEE+WL&$\Lambda$CDM&$\Lambda$CDM+$A_{\rm lens}$\\
\hline\morehorsp
$\Lambda$CDM (HZ)& $19.17$ & $18.55$\\ 
$\Lambda$CDM+$A_{\rm lens}$ (HZ)& $14.11$ & $13.49$\\ 
\hline
\end{tabular}
\end{center}
\caption{Bayesian evidence for an HZ spectrum in case of $\Lambda$CDM and $\Lambda$CDM+$A_{\rm lens}$. Planck TT+WL and Planck TTTEEE+WL data are considered.}
\label{tablenew4}
\end{table*}

\subsection{Planck + WL}

As discussed in the introduction the Planck dataset shows an internal tension above the $2$ standard deviations on the determination of the amplitude of the lensing parameter $A_{\rm lens}$. Interestingly, the inclusion of $A_{\rm lens}$ as a free parameter in the Planck analysis results in a $\sigma_8$ estimate that is in better agreement with the one obtained from cosmic shear surveys. It is therefore important to assess the viability of an HZ model in the framework of a $\Lambda$CDM+$A_{\rm lens}$ model when considering cosmic shear data -  we use the revised version of the CFHTLenS cosmic shear dataset \cite{planck2015}.  

In Table~\ref{tablenew4} we report the Bayesian evidence for the Planck+WL dataset, including the possibility of a variation in $A_{\rm lens}$. We see very strong evidence against an HZ spectrum in most cases. However, if we limit just to Planck TT +WL, the evidence against HZ in a $\Lambda$CDM+$A_{\rm lens}$ scenario is just marginal when compared either with $\Lambda$CDM, either with $\Lambda$CDM+$A_{\rm lens}$ itself.

Interestingly, including the polarization data changes this conclusion quite dramatically.
Indeed, if we now focus attention on the results in the last row of Table~\ref{tablenew4}, we see that an HZ spectrum in a $\Lambda$CDM+$A_{\rm lens}$ framework is strongly ruled out by the Planck TTTEEE+WL dataset with a very strong negative evidence ($\Delta \ln {\cal B} > 5$).

In summary, the $A_{\rm lens}$ tension brings an HZ spectrum back in to agreement with Planck TT data but not when the Planck TTTEEE data are considered.

\section{Conclusions}

In this paper we have discussed the agreement of a Harrison-Zel'dovich primordial power spectrum with  cosmological data under the assumption of extended cosmological scenarios motivated by tensions between current cosmological datasets. This is an important analysis since having very strong evidence against HZ even in extended scenarios would further support inflation.

As already pointed out in the literature, we have shown that an HZ spectrum, in the framework of $\Lambda$CDM, is indeed strongly disfavored by Planck temperature and polarization data with very strong evidence against it.

However, focusing just on Planck TT data, we have found no significant evidence against HZ when considering variations in the neutrino number $N_{\rm eff}$, in the Helium abundance $Y_{\rm He}$ and in a combination of the two. Furthermore we have found even a positive evidence for HZ with respect to $\Lambda$CDM when R16 is included.

The Planck TT result changes with the inclusion of polarization data, which improves the determination of $N_{\rm eff}$, producing now from Planck TTTEEE data strong evidence against HZ with respect to $\Lambda$CDM+$N_{\rm eff}$ and very strong evidence against HZ within $\Lambda$CDM. 

This is mitigated by the inclusion of R16 data. From the Planck TTTEEE+R16 dataset we found only positive evidence against HZ with respect to $\Lambda$CDM and inconclusive evidence with respect to $\Lambda$CDM+$N_{\rm eff}$ and $\Lambda$CDM+$Y_{\rm He}$. 

If we include information from BAO, we have found very strong evidence against HZ in all cases with the exception of the $\Lambda$CDM+$N_{\rm eff}$ scenario.

Therefore, when considering the $\Lambda$CDM+$N_{\rm eff}$ scenario we can state that R16 and BAO data have opposite effects in ruling out HZ. R16 is in someway reducing the discrepancy with HZ while BAO data increases it.

If we include information from cosmic shear, we have found from Planck TT data very strong evidence against HZ assuming $\Lambda$CDM but no significant evidence against HZ in the case of a $\Lambda$CDM+$A_{\rm lens}$ scenario. However, the inclusion of Planck polarization data again works against HZ and we found very strong evidence against HZ from Planck TTTEEE+WL data even when allowing $A_{\rm lens}$ to vary. 

We have also investigated if further parameter extensions could alter the conclusions. When polarization data are included, there is always a very strong evidence against these extensions with respect to $\Lambda$CDM due to the increased number of parameters, but within these extended parameter frameworks, an HZ spectrum is not yet ruled out, with strong evidence in favor of it when considering the Planck TTTEEE+R16 dataset and the Extended-$10$ scenario.

The possibility of fully ruling out the HZ spectrum with very strong evidence is therefore ultimately connected with the solution to the current tension on the Hubble parameter between Planck and R16.
If the tension is confirmed by future data, then new physical mechanisms could be at work and an HZ spectrum could still offer a possible alternative. 

\
\section*{Acknowledgements}
EDV acknowledges support from the European Research Council in the form of a Consolidator Grant with number 681431. YF is supported by the Robert Bosch Stiftung. AM thanks the University of Manchester and the Jodrell Bank Center for Astrophysics for hospitality. AM acknowledges support from TASP, iniziativa specifica INFN.

\section{Appendix}

In this appendix we discuss the constraints on cosmological parameters from the several analyses performed.

\begin{table*}
\begin{center}\footnotesize
\scalebox{0.74}{\begin{tabular}{c|cccccccc}
\hline
\hspace{1mm}\\
Parameter         & $\Lambda$CDM & $\Lambda$CDM (HZ) & $\Lambda$CDM +$N_{\rm eff}$ & $\Lambda$CDM +$N_{\rm eff}$ (HZ) & Extended-10 & Extended-10 (HZ) & Extended-11 & Extended-11 (HZ)   \\       
\hline
\hspace{1mm}\\
$\Omega_{\rm b}h^2$  &  $0.02222\,\pm 0.00023$& $0.02300\,\pm 0.00020$  & $0.02230\,\pm 0.00037$& $0.02294\, \pm 0.00019$ &  $0.02296\,^{+0.00062}_{-0.00075}$& $0.02304\,\pm 0.00028$  & $0.02296\, ^{+0.00067}_{-0.00083}$& $0.02304\,\pm 0.00029$ \\
\hspace{1mm}\\
$\Omega_{\rm c}h^2$  &  $0.1198\,\pm 0.0022$&  $0.1100\,\pm 0.0011$&  $0.1205\,\pm 0.0041$&  $0.1248\,\pm 0.0034$ &  $0.1220\,^{+0.0062}_{-0.0078} $&  $0.1224\,^{+0.0037}_{-0.0044}$&  $0.1216\,_{-0.0075}^{+0.0061}$&  $0.1222\,\pm 0.0042$\\
\hspace{1mm}\\
$\theta_{\rm c}$  &  $1.04085\,\pm 0.00048$&  $1.04217\,\pm 0.00041$&  $1.04082\,\pm 0.00056$&  $1.04055\,\pm 0.00052$ &  $1.04073\,\pm 0.00076 $&  $1.04065\,\pm 0.00065$&  $1.04078\,^{+0.00073}_{-0.00083}$&  $1.04066\,\pm 0.00067$\\
\hspace{1mm}\\
$\tau$ &   $0.077\,\pm 0.019$& $0.139\,^{+0.019}_{-0.017}$&  $0.080\,\pm 0.022$& $0.110\, \pm 0.019$ &   $0.067\, \pm 0.024$& $0.068\,^{+0.023}_{-0.026}$&  $0.067\, \pm 0.024$& $0.066\, \pm 0.023$ \\
\hspace{1mm}\\
$n_{\rm s}$ &  $0.9655\,\pm 0.0062$&$1$  & $0.969\,\pm 0.016$&$1$ &  $0.995\,^{+0.033}_{-0.038}$&$1$  & $0.994\,_{-0.040}^{+0.035} $&$1$  \\
\hspace{1mm}\\
$\ln(10^{10}A_{\rm s})$ & $3.088\,\pm 0.036$& $3.189\,^{+0.039}_{-0.033}$& $3.096\,\pm 0.047$& $3.166\,^{+0.039}_{-0.035}$ & $3.070\,\pm 0.050$& $3.074\,^{+0.046}_{-0.052}$& $3.069\, \pm 0.050$& $3.070\,\pm 0.047$\\
\hspace{1mm}\\
$H_0 /\rm{km \, s^{-1} \, Mpc^{-1}}$ &$67.29\,\pm 0.98$ &$72.01\,\pm 0.51$&$68.0\,\pm 2.8$ &$73.51\,\pm 0.64$ &$69.6\,^{+5.9}_{-8.2}$ &$70.1\,^{+3.7}_{-2.2}$&$70\, ^{+10}_{-20}$ &$70\,^{+10}_{-20}$  \\
\hspace{1mm}\\
$\sigma_8$ &$0.829\,\pm 0.014$& $0.842\,\pm 0.016$  & $0.834\,\pm 0.023$& $0.868\,\pm 0.017$ &$0.733\,_{-0.084}^{+0.064}$& $0.734\,^{+0.087}_{-0.055}$  & $0.72\,_{-0.12}^{+0.18}$& $0.71\,^{+0.11}_{-0.18}$ \\
\hspace{1mm}\\
$\neff$ & $3.046$ &$3.046$& $3.12\,\pm 0.31$ &$3.69\,\pm 0.14$& $3.59\,^{+0.61}_{-0.85}$ &$3.64\,^{+0.16}_{-0.19}$ & $3.55\,^{+0.62}_{-0.85}$ &$2.63\,^{+0.16}_{-0.20}$ \\
\hspace{1mm}\\
$ \Sigma m_\nu[\, eV]$ &              $0.06$&$0.06$&$0.06$&$0.06$&$<0.627$&$<0.628$&$<0.631$&$<0.668$ \\
\hspace{1mm}\\
$ d\ln n_{\rm s}/d\ln k$ & $0$&$0$&$0$&$0$& $0.006\,\pm0.016$&$0.0078\,\pm0.0095$& $0.004\,^{+0.015}_{-0.017}$&$0.007\,\pm 0.010$ \\
\hspace{1mm}\\
$ A_{\rm lens}$ &  $1$&$1$&$1$&$1$& $1.35\,^{+0.15}_{-0.18}$&$1.37\,^{+0.11}_{-0.13}$& $1.43\,^{+0.16}_{-0.31}$&$1.45\,^{+0.14}_{-0.22}$ \\
\hspace{1mm}\\
$ w$ &   $-1$&$-1$&$-1$&$-1$&$-1$&$-1$&$>-1.27$&$>-1.22$ \\
\hline
\hspace{1mm}\\
$\bar{\chi}^2_{\rm eff}$ & $11281.95$&$11307.88$&$11282.90$&$11286.19$ &$11279.57$&$11278.64$&$11279.30$&$11278.39$ \\
\hline
\end{tabular}}
\end{center}
\label{tab:TT}
\caption{$68\%$ credible intervals for cosmological parameters for the Planck TT dataset and for several cosmological frameworks. The $\bar{\chi}^2_{\rm eff}$ reported are from the Planck TT dataset.}
\end{table*}

\subsection{Planck TT}

Here we report the constraints on cosmological parameters from the Planck TT dataset under the assumption of $\Lambda$CDM, $\Lambda$CDM+$N_{\rm eff}$, Extended-$10$, and Extended-$11$ models in Table~\ref{tab:TT}. The results for the $\Lambda$CDM+$Y_{\rm He}$ and $\Lambda$CDM+$Y_{\rm He}$+$N_{\rm eff}$ models using the same data set are presented in the first columns of Table~\ref{tab:yhe} and ~\ref{tab:yheneff}.  Although the main conclusions come from the Bayesian Evidence, for completeness we report the mean effective chi-square, $\bar{\chi}^2_{\rm eff}$, computed by weighting the $\chi^2$ values of the models present in the MCMC chains, at the bottom of each Table. This quantity can give an idea, albeit not fully rigorous, of the goodness-of-fit of the selected scenario (see \cite{planck2015}). As we can see, in the case of standard $\Lambda$CDM, the HZ spectrum is strongly disfavored with $\Delta \bar{\chi}^2_{\rm eff} \sim 26$. We also note that the assumption of HZ introduces a major shift in most of the parameters. In particular the $\Lambda$CDM HZ model prefers a higher value for the optical depth $\tau$, a higher Hubble constant of $H_0=72.01\pm0.51$ km s$^{-1}$ Mpc$^{-1}$ at $68 \%$, i.e. in agreement with the R16 constraint \cite{R16}, a smaller value for the cold dark matter density and a higher value for the baryon density.

When we move to the case of $\Lambda$CDM+$N_{\rm eff}$ we see that the introduction of $N_{\rm eff}$ essentially weakens the constraints on the Hubble constant by nearly a factor $3$ and the constraints on the baryon and cold dark matter densities and $n_{\rm s}$ by nearly a factor of two. The mean values of the parameters are almost the same as in the case of $\Lambda$CDM. A variation in $N_{\rm eff}$ changes the epochs of equality and decoupling affecting the sound horizon scale $r_s$ and the silk damping scale $r_d$. Moreover varying $N_{\rm eff}$ introduces the possibility of changing the early integrated Sachs-Wolfe effect that shifts the peaks positions and is degenerate with $\theta_{\rm c}$. This introduces a further degeneracy between the parameters that explains the weakening of the constraints. The value of $\bar{\chi}^2_{\rm eff}$ is practically unchanged from $\Lambda$CDM. When the HZ spectrum is assumed in the $\Lambda$CDM+$N_{\rm eff}$ scenario we note first a strong indication for $N_{\rm eff}>3.046$. Indeed, the $n_{\rm s}=1$ spectrum, with pivot scale at $k_{\rm p}=0.05 hMpc^{-1}$, shows a CMB first peak in the TT spectrum that is lower with respect to the $n_{\rm s}=0.969$ model. Increasing $N_{\rm eff}$ adds power to the first peak owing to the early integrated Sachs-Wolfe effect and helps in reconciling HZ with data. We also see that the Hubble constant is again increased and in perfect agreement with R16. The cold dark matter density, however, is larger with respect to the varying $n_{\rm s}$ case. The assumption of HZ results in a moderate increase of $\Delta \bar{\chi}^2_{\rm eff}\sim 3.5$, i.e. 
HZ is in better agreement with the data when variable $N_{\rm eff}$ is considered. If instead of $N_{\rm eff}$ we vary $Y_{\rm He}$, we see from the constraints in Table~\ref{tab:yhe} that the bounds on $n_{\rm s}$ are strongly weakened - increasing the error by almost a factor $2$ with respect to $\Lambda$CDM. Variation on $Y_{\rm He}$ has a smaller effect on the constraint of the Hubble constant. In summary, variable $N_{\rm eff}$ and variable $Y_{\rm He}$ both weaken the constraints on $n_{\rm s}$, but only $N_{\rm eff}$ significantly shifts the mean value and broadens the constraint on $H_0$.
When both $N_{\rm eff}$ and $Y_{\rm He}$ are varied (see Table ~\ref{tab:yheneff}) the constraints on $n_{\rm s}$ are further enlarged by $\sim 60 \%$. As in the $\Lambda$CDM+$N_{\rm eff}$ case, there is a moderate increment in the value of the $\bar{\chi}^2_{\rm eff}$ when imposing an HZ spectrum, with $\Delta \bar{\chi}^2_{\rm eff} \sim 4-5$,

When we consider the Extended-$10$ and Extended-$11$ scenarios as shown in the Table~\ref{tab:TT}, we note the following: a) the $\bar{\chi}^2_{\rm eff}$ is always very close and slightly better than the $\Lambda$CDM case; b) There is very little variation in $\bar{\chi}^2_{\rm eff}$ when HZ is introduced, i.e. HZ can't be ruled out on the basis of a simple $\bar{\chi}^2_{\rm eff}$ analysis in these extended scenarios; c) The main effect of assuming an HZ is to provide evidence for $N_{\rm eff}>3.046$ and to further improve the indication for $A_{\rm lens}>1$; d) The Hubble constant is left practically as undetermined even when imposing HZ.

\begin{table*}
\begin{center}\footnotesize
\scalebox{0.74}{\begin{tabular}{c|cccccccc}
\hline
\hspace{1mm}\\
Parameter         & $\Lambda$CDM & $\Lambda$CDM (HZ) & $\Lambda$CDM +$N_{\rm eff}$ & $\Lambda$CDM +$N_{\rm eff}$ (HZ) & Extended-10 & Extended-10 (HZ) & Extended-11 & Extended-11 (HZ)   \\       
\hline
\hspace{1mm}\\
$\Omega_{\rm b}h^2$  &  $0.02247\,\pm 0.00022$& $0.02302\,\pm 0.00019$  & $0.02275\,\pm 0.00025$& $0.02295\, \pm 0.00019$ &  $0.02326\,\pm 0.00031$& $0.02315\,\pm 0.00026$  & $0.02286\, ^{+0.00061}_{-0.00078}$& $0.02301\,\pm 0.00029$ \\
\hspace{1mm}\\
$\Omega_{\rm c}h^2$  &  $0.1167\,\pm 0.0019$&  $0.1098\,\pm 0.0011$&  $0.1236\,\pm 0.0036$&  $0.1251\,\pm 0.0034$ &  $0.1239\,^{+0.0045}_{-0.0059} $&  $0.1215\,\pm 0.0038$&  $0.1215\,_{-0.0073}^{+0.0061}$&  $0.1227\,\pm 0.0040$\\
\hspace{1mm}\\
$\theta_{\rm c}$  &  $1.04130\,\pm 0.00045$&  $1.04221\,\pm 0.00041$&  $1.04062\,\pm 0.00054$&  $1.04050\,\pm 0.00052$ &  $1.04061\,^{+0.00075}_{-0.00068} $&  $1.04087\,\pm 0.00059$&  $1.04074\,\pm 0.00076$&  $1.04060\,\pm 0.00067$\\
\hspace{1mm}\\
$\tau$ &   $0.091\,\pm 0.020$& $0.140\,^{+0.019}_{-0.017}$&  $0.099\,\pm 0.020$& $0.109\, \pm 0.018$ &   $0.068\, \pm 0.024$& $0.068\,\pm 0.024$&  $0.064\, \pm 0.023$& $0.067\, \pm 0.023$ \\
\hspace{1mm}\\
$n_{\rm s}$ &  $0.9729\,\pm 0.0057$&$1$  & $0.9899\,^{+0.0095}_{-0.0086}$&$1$ &  $1.011\,^{+0.015}_{-0.017}$&$1$  & $0.991\,_{-0.038}^{+0.033} $&$1$  \\
\hspace{1mm}\\
$\ln(10^{10}A_{\rm s})$ & $3.109\,\pm 0.038$& $3.190\,^{+0.038}_{-0.034}$& $3.143\,\pm 0.040$& $3.165\,\pm 0.035$ & $3.077\,\pm 0.048$& $3.075\,\pm 0.048$& $3.063\, \pm 0.050$& $3.072\,\pm 0.047$\\
\hspace{1mm}\\
$H_0 /\rm{km \, s^{-1} \, Mpc^{-1}}$ &$68.73\,\pm 0.86$ &$72.10\,\pm 0.50$&$71.9\,\pm 1.6$ &$73.52\,\pm 0.61$ &$72.8\,\pm 1.7$ &$72.4\,^{+1.6}_{-1.0}$&$73.6\, \pm 2.0$ &$73.4\,\pm 1.9$  \\
\hspace{1mm}\\
$\sigma_8$ &$0.829\,\pm 0.015$& $0.842\,\pm 0.016$  & $0.856\,\pm 0.019$& $0.868\,\pm 0.017$ &$0.756\,_{-0.044}^{+0.071}$& $0.789\,^{+0.043}_{-0.033}$  & $0.763\,\pm 0.064$& $0.755\,^{+0.059}_{-0.049}$ \\
\hspace{1mm}\\
$\neff$ & $3.046$ &$3.046$& $3.52\,\pm 0.20$ &$3.70\,\pm 0.14$& $3.85\,^{+0.30}_{-0.48}$ &$3.59\,\pm 0.15$ & $3.51\,^{+0.60}_{-0.82}$ &$3.65\,^{+0.16}_{-0.19}$ \\
\hspace{1mm}\\
$ \Sigma m_\nu[\, eV]$ &              $0.06$&$0.06$&$0.06$&$0.06$&$<0.506$&$<0.267$&$<0.659$&$<0.677$ \\
\hspace{1mm}\\
$ d\ln n_{\rm s}/d\ln k$ & $0$&$0$&$0$&$0$& $0.011\,^{+0.011}_{-0.014}$&$0.0043\,\pm0.0082$& $0.006\,\pm 0.015$&$0.0088\,\pm 0.0097$ \\
\hspace{1mm}\\
$ A_{\rm lens}$ &  $1$&$1$&$1$&$1$& $1.39\,^{+0.13}_{-0.16}$&$1.32\,^{+0.10}_{-0.12}$& $1.34\,^{+0.15}_{-0.19}$&$1.36\,^{+0.11}_{-0.13}$ \\
\hspace{1mm}\\
$ w$ &   $-1$&$-1$&$-1$&$-1$&$-1$&$-1$& $-1.22\,^{+0.36}_{-0.22}$&$-1.12\,^{+0.16}_{-0.09}$ \\
\hline
\hspace{1mm}\\
$\bar{\chi}^2_{\rm eff}$ & $11290.52$&$11308.44$&$11286.00$&$11286.32$ &$11279.96$&$11278.49$&$11280.44$&$11279.64$ \\
\hline
\end{tabular}}
\end{center}
\caption{$68\%$ credible intervals for cosmological parameters for the Planck TT + R16 dataset and for several cosmological frameworks. The $\bar{\chi}^2_{\rm eff}$ reported are from the Planck TT + R16 dataset.
Note that now an HZ spectrum produces similar $\bar{\chi}^2_{\rm eff}$ in the case of $\Lambda$CDM+$N_{\rm eff}$.}
\label{tab:TT+R16}
\end{table*}

\subsection{Planck TT+R16}

In Table~\ref{tab:TT+R16},~\ref{tab:yhe}, and Table~\ref{tab:yheneff} we report the  $68 \%$ credible intervals for the cosmological parameters using the Planck TT + R16 dataset. We can immediately see that the inclusion of R16 forces the spectral index $n_{\rm s}$ to be slightly higher in the case of $\Lambda$CDM. Not surprisingly, the $\bar{\chi}^2_{\rm eff}$ is definitely worse  for $\Lambda$CDM in the case of Planck TT +R16, as a result of the tension between the Planck data and R16. However we see that this is not the case when comparing the $\bar{\chi}^2_{\rm eff}$ values for models with an HZ spectrum between Planck TT and Planck TT+R16.
As we discussed in the previous paragraph, the assumption of the HZ spectrum shifts the values of the Hubble constant in agreement with R16. Therefore, for these models, the inclusion of R16 has little impact on the $\bar{\chi}^2_{\rm eff}$. Imposing in the case of the Planck TT+R16 dataset an HZ spectrum in $\Lambda$CDM increases the $\bar{\chi}^2_{\rm eff}$ by $\sim 18$, i.e. a smaller value respect to the Planck TT case.

As we can see, now the $\Lambda$CDM+$N_{\rm eff}$ scenario produces a fit to Planck TT+R16 that is better than the one achievable assuming $\Lambda$CDM, i.e. the inclusion of $N_{\rm eff}$ helps in solving the tension on the Hubble constant. With respect to Planck TT alone, the inclusion of R16 increases the effective mean chi-square by just $\Delta \bar{\chi}^2_{\rm eff}\sim 3$. More importantly, we can also see that imposing HZ in this scenario does not worsen the $\bar{\chi}^2_{\rm eff}$, i.e. HZ is now fully consistent with the data. 

Looking at Table~\ref{tab:yhe} and ~\ref{tab:yheneff}, an increase of $\Delta \bar{\chi}^2_{\rm eff} \sim 7$ is however present when including R16 in case of $\Lambda$CDM+$Y_{\rm He}$ and $\Lambda$CDM+$Y_{\rm He}$+$N_{\rm eff}$
 with respect to Planck TT alone. Indeed, as we discussed in the previous paragraph, including $Y_{\rm He}$ weakens the bounds on $n_{\rm s}$ but less significantly on $H_0$, i.e. it does not fully help in solving the Hubble tension. However, in both $\Lambda$CDM+$Y_{\rm He}$ and $\Lambda$CDM+$Y_{\rm He}$+$N_{\rm eff}$ imposing HZ has negligible effect on $\bar{\chi}^2_{\rm eff}$.

If we look at the small differences in the $\bar{\chi}^2_{\rm eff}$ values  in Table~\ref{tab:TT+R16} we can conclude that HZ is also consistent with Planck TT+R16 when we consider the Extended-$10$ and Extended-$11$ models.

\begin{table*}
\begin{center}\footnotesize
\scalebox{0.74}{\begin{tabular}{c|cccccccc}
\hline
\hspace{1mm}\\
Parameter         & $\Lambda$CDM & $\Lambda$CDM (HZ) & $\Lambda$CDM +$N_{\rm eff}$ & $\Lambda$CDM +$N_{\rm eff}$ (HZ) & Extended-10 & Extended-10 (HZ) & Extended-11 & Extended-11 (HZ)   \\       
\hline
\hspace{1mm}\\
$\Omega_{\rm b}h^2$  &  $0.02226\,\pm 0.00015$& $0.02285\,\pm 0.00014$  & $0.02219\,\pm 0.00025$& $0.02298\, \pm 0.00014$ &  $0.02227\,\pm 0.00028$& $0.02295\,\pm 0.00016$  & $0.02225\, \pm 0.00028$& $0.02295\,\pm 0.00016$ \\
\hspace{1mm}\\
$\Omega_{\rm c}h^2$  &  $0.1198\,\pm 0.0014$&  $0.11166\,\pm 0.00087$&  $0.1189\,\pm 0.0031$&  $0.1262\,\pm 0.0026$ &  $0.1186\,\pm 0.0034 $&  $0.1253\,\pm 0.0028$&  $0.1186\,\pm 0.0034$&  $0.1253\,\pm 0.0029$\\
\hspace{1mm}\\
$\theta_{\rm c}$  &  $1.04077\,\pm 0.00032$&  $1.04171\,\pm 0.00029$&  $1.04088\,\pm 0.00044$&  $1.04016\,\pm 0.00038$ &  $1.04073\,\pm 0.00051 $&  $1.04005\,\pm 0.00043$&  $1.04071\,\pm 0.00052$&  $1.04004\,\pm 0.00044$\\
\hspace{1mm}\\
$\tau$ &   $0.079\,\pm 0.017$& $0.143\,\pm 0.016$&  $0.077\,\pm 0.018$& $0.114\, \pm 0.016$ &   $0.059\, \pm 0.021$& $0.061\,\pm 0.022$&  $0.058\, \pm 0.021$& $0.061\, \pm 0.021$ \\
\hspace{1mm}\\
$n_{\rm s}$ &  $0.9646\,\pm 0.0047$&$1$  & $0.9618\,\pm 0.0099$&$1$ &  $0.964\,\pm 0.013$&$1$  & $0.964\,\pm 0.012 $&$1$  \\
\hspace{1mm}\\
$\ln(10^{10}A_{\rm s})$ & $3.094\,\pm 0.034$& $3.199\,\pm 0.032$& $3.087\,\pm 0.038$& $3.177\,\pm 0.031$ & $3.049\,\pm 0.044$& $3.065\,\pm 0.044$& $3.046\, ^{+0.043}_{-0.048}$& $3.064\,\pm 0.042$\\
\hspace{1mm}\\
$H_0 /\rm{km \, s^{-1} \, Mpc^{-1}}$ &$67.30\,\pm 0.64$ &$71.07\,\pm 0.42$&$66.8\,\pm 1.6$ &$73.00\,\pm 0.56$ &$63.9\,\pm 3.0$ &$69.6\,^{+3.2}_{-2.2}$&$74\, \pm 10$ &$73\,\pm 20$  \\
\hspace{1mm}\\
$\sigma_8$ &$0.831\,^{+0.015}_{-0.013}$& $0.854\,\pm 0.014$  & $0.827\,^{+0.017}_{-0.020}$& $0.877\,\pm 0.014$ &$0.722\,_{-0.060}^{+0.076}$& $0.740\,^{+0.078}_{-0.057}$  & $0.79\,_{-0.14}^{+0.16}$& $0.75\,\pm 0.13$ \\
\hspace{1mm}\\
$\neff$ & $3.046$ &$3.046$& $2.98\,\pm 0.20$ &$3.70\,\pm 0.11$& $3.03\,\pm 0.25$ &$3.71\,^{+0.11}_{-0.14}$ & $3.03\,\pm 0.25$ &$3.71\,^{+0.12}_{-0.14}$ \\
\hspace{1mm}\\
$ \Sigma m_\nu[\, eV]$ & $0.06$&$0.06$&$0.06$&$0.06$&$<0.606$ &$0.51\,^{+0.13}_{-0.50}$&$0.53\,^{+0.21}_{-0.45}$&$0.55\,^{+0.18}_{-0.50}$ \\
\hspace{1mm}\\
$ d\ln n_{\rm s}/d\ln k$ & $0$&$0$&$0$&$0$& $-0.0014\,\pm0.0087$&$0.0137\,\pm0.0074$& $-0.0005\,\pm 0.0088$&$0.0138\,\pm 0.0079$ \\
\hspace{1mm}\\
$ A_{\rm lens}$ &  $1$&$1$&$1$&$1$& $1.22\,^{+0.10}_{-0.12}$&$1.33\,^{+0.10}_{-0.12}$& $1.22\,^{+0.10}_{-0.14}$&$1.37\,^{+0.11}_{-0.17}$ \\
\hspace{1mm}\\
$ w$ &   $-1$&$-1$&$-1$&$-1$&$-1$&$-1$& $-1.39\,\pm0.58$&$>-1.40$ \\
\hline
\hspace{1mm}\\
$\bar{\chi}^2_{\rm eff}$ & $12967.40$&$13016.58$&$12968.37$&$12982.53$ &$12965.74$&$12973.03$&$12965.95$&$12973.25$ \\
\hline
\end{tabular}}
\end{center}
\caption{$68\%$ credible intervals for cosmological parameters for the Planck TTTEEE dataset and for several cosmological frameworks. The $\bar{\chi}^2_{\rm eff}$ reported are from the Planck TTTEEE dataset.}
\label{tab:TTTEEE}
\end{table*}

\subsection{Planck TTTEEE}

As we can see in Table~\ref{tab:TTTEEE}, Planck polarization data significantly improves the constraints on cosmological parameters. 
For example, if we focus attention on the simple $\Lambda$CDM model we see that the inclusion of CMB polarization data increases the accuracy on $n_{\rm s}$ by $\sim 25 \%$. The consequence of this is that now an HZ spectrum in the $\Lambda$CDM scenario is ruled out even more with $\Delta \bar{\chi}^2_{\rm eff} \sim 39$.
The interesting point is that an HZ spectrum is significantly disfavored also when considering the inclusion of $N_{\rm eff}$. The assumption of HZ produces a worse fit to the data with $\Delta \bar{\chi}^2_{\rm eff} \sim 14$ in the $\Lambda$CDM+$N_{\rm eff}$ scenario. Polarization data indeed increases significantly the constraint on $N_{\rm eff}$ by more than $30 \%$ with a mean value close to the standard expectation of $N_{\rm eff}=3.046$. The physical reason for this is that polarization data are unaffected by the additional early ISW produced by a larger $N_{\rm eff}$. Including polarization therefore helps in its determination and breaks some of the degeneracies between, for example, $N_{\rm eff}$ and $\Omega_{\rm c}h^2$. A similar argument is also valid in case of $Y_{\rm He}$. As we can see from Table~\ref{tab:yhe} and ~\ref{tab:yheneff}, the inclusion of polarization data significantly improves the constraints on $n_{\rm s}$ also in these cases.

\begin{table*}
\begin{center}\footnotesize
\scalebox{0.74}{\begin{tabular}{c|cccccccc}
\hline
\hspace{1mm}\\
Parameter         & $\Lambda$CDM & $\Lambda$CDM (HZ) & $\Lambda$CDM +$N_{\rm eff}$ & $\Lambda$CDM +$N_{\rm eff}$ (HZ) & Extended-10 & Extended-10 (HZ) & Extended-11 & Extended-11 (HZ)   \\       
\hline
\hspace{1mm}\\
$\Omega_{\rm b}h^2$  &  $0.02236\,\pm 0.00016$& $0.02287\,\pm 0.00014$  & $0.02258\,\pm 0.00019$& $0.02297\, \pm 0.00014$ &  $0.02278\,\pm 0.00022$& $0.02301\,\pm 0.00016$  & $0.02225\, \pm 0.00028$& $0.02295\,\pm 0.00016$ \\
\hspace{1mm}\\
$\Omega_{\rm c}h^2$  &  $0.1184\,\pm 0.0014$&  $0.11142\,\pm 0.00085$&  $0.1228\,\pm 0.0029$&  $0.1262\,\pm 0.0026$ &  $0.1222\,\pm 0.0031 $&  $0.1247\,\pm 0.0027$&  $0.1187\,\pm 0.0034$&  $0.1252\,\pm 0.0028$\\
\hspace{1mm}\\
$\theta_{\rm c}$  &  $1.04094\,\pm 0.00032$&  $1.04173\,\pm 0.00029$&  $1.04049\,\pm 0.00040$&  $1.04017\,\pm 0.00037$ &  $1.04051\,\pm 0.00044 $&  $1.04023\,\pm 0.00040$&  $1.04070\,\pm 0.00051$&  $1.04004\,\pm 0.00043$\\
\hspace{1mm}\\
$\tau$ &   $0.086\,\pm 0.017$& $0.143\,^{+0.017}_{-0.015}$&  $0.092\,\pm 0.017$& $0.114\, \pm 0.016$ &   $0.059\, \pm 0.021$& $0.061\,\pm 0.021$&  $0.059\, \pm 0.021$& $0.061\, \pm 0.022$ \\
\hspace{1mm}\\
$n_{\rm s}$ &  $0.9682\,\pm 0.0048$&$1$  & $0.9787\,\pm 0.0077$&$1$ &  $0.9857\,\pm 0.0092$&$1$  & $0.964\,\pm 0.013 $&$1$  \\
\hspace{1mm}\\
$\ln(10^{10}A_{\rm s})$ & $3.105\,\pm 0.033$& $3.199\,^{+0.035}_{-0.029}$& $3.127\,\pm 0.035$& $3.178\,\pm 0.031$ & $3.057\,\pm 0.043$& $3.065\,\pm 0.043$& $3.047\, ^{+0.044}_{-0.049}$& $3.065\,\pm 0.045$\\
\hspace{1mm}\\
$H_0 /\rm{km \, s^{-1} \, Mpc^{-1}}$ &$67.91\,\pm 0.64$ &$71.19\,\pm 0.41$&$69.8\,\pm 1.3$ &$72.98\,\pm 0.53$ &$70.5\,\pm 1.4$ &$72.1\,^{+1.5}_{-1.0}$&$73.9\, \pm 2.0$ &$73.5\,\pm 1.9$  \\
\hspace{1mm}\\
$\sigma_8$ &$0.832\,\pm 0.013$& $0.853\,^{+0.015}_{-0.013}$  & $0.849\,\pm 0.017$& $0.877\,\pm 0.014$ &$0.806\,_{-0.024}^{+0.033}$& $0.799\,^{+0.041}_{-0.030}$  & $0.797\,\pm 0.053$& $0.767\,^{+0.060}_{-0.047}$ \\
\hspace{1mm}\\
$\neff$ & $3.046$ &$3.046$& $3.34\,\pm 0.17$ &$3.70\,\pm 0.11$& $3.41\,^{+0.19}_{-0.21}$ &$3.67\,\pm 0.11$ & $3.03\,\pm 0.25$ &$3.70\,^{+0.12}_{-0.14}$ \\
\hspace{1mm}\\
$ \Sigma m_\nu[\, eV]$ & $0.06$&$0.06$&$0.06$&$0.06$&$<0.149$ &$<0.244$&$0.52\,^{+0.21}_{-0.44}$&$0.53\,^{+0.16}_{-0.51}$ \\
\hspace{1mm}\\
$ d\ln n_{\rm s}/d\ln k$ & $0$&$0$&$0$&$0$& $0.0050\,\pm0.0078$&$0.0112\,\pm0.0068$& $0.0004\,\pm 0.0086$&$0.0144\,\pm 0.0076$ \\
\hspace{1mm}\\
$ A_{\rm lens}$ &  $1$&$1$&$1$&$1$& $1.220\,^{+0.085}_{-0.098}$&$1.277\,^{+0.083}_{-0.098}$& $1.20\,\pm 0.10$&$1.32\,^{+0.10}_{-0.11}$ \\
\hspace{1mm}\\
$ w$ &   $-1$&$-1$&$-1$&$-1$&$-1$&$-1$& $-1.42\,^{+0.25}_{-0.15}$&$-1.15\,^{+0.15}_{-0.09}$ \\
\hline
\hspace{1mm}\\
$\bar{\chi}^2_{\rm eff}$ & $12977.06$&$13018.21$&$12975.27$&$12982.42$ &$12971.20$&$12973.47$&$12966.78$&$12974.11$ \\
\hline
\end{tabular}}
\end{center}
\caption{$68\%$ credible intervals for cosmological parameters for the Planck TTTEEE +R16 dataset and for several cosmological frameworks.
The $\bar{\chi}^2_{\rm eff}$ reported are from the Planck TTTEEE+R16 dataset.}
\label{tab:TTTEEE+R16}
\end{table*}

\begin{table*}
\begin{center}\footnotesize
\scalebox{0.74}{\begin{tabular}{c|cccccccc}
\hline
Parameter         & $\Lambda$CDM+$Y_{\rm He}$ & $\Lambda$CDM+$Y_{\rm He}$ (HZ) & $\Lambda$CDM +$Y_{\rm He}$ & $\Lambda$CDM +$Y_{\rm He}$ (HZ) & $\Lambda$CDM+$Y_{\rm He}$ & $\Lambda$CDM+$Y_{\rm He}$ (HZ) & $\Lambda$CDM +$Y_{\rm He}$ & $\Lambda$CDM +$Y_{\rm He}$ (HZ)   \\  
\hspace{1mm}\\
       & TT & TT & TT+R16 & TT+R16 & TTTEEE & TTTEEE & TTTEEE+R16 & TTTEEE+R16  \\ 
\hline
\hspace{1mm}\\
$\Omega_{\rm b}h^2$  &  $0.02229\,\pm 0.00033$& $0.02302\,\pm 0.00020$  & $0.02276\,\pm 0.00031$& $0.02306\, \pm 0.00019$ &  $0.02230\,\pm 0.00022$& $0.02301,\pm 0.00014$  & $0.02251\,\pm 0.00022$& $0.02305\, \pm 0.00014$\\
\hspace{1mm}\\
$\Omega_{\rm c}h^2$  &  $0.1196\,\pm 0.0024$&  $0.1153\,\pm 0.0016$&  $0.1160\,\pm 0.0020$&  $0.1145\,\pm 0.0015$ &  $0.1198\,\pm 0.0015$&  $0.1162\,\pm 0.0012$&  $0.1181\,\pm 0.0014$&  $0.1157\,\pm 0.0011$ \\
\hspace{1mm}\\
$\theta_{\rm c}$  &  $1.04107\,\pm 0.00093$&  $1.04317\,\pm 0.00047$&  $1.04231\,\pm 0.00085$&  $1.04319\,\pm 0.00046$ &  $1.04095\,\pm 0.00059$&  $1.04287\,\pm 0.00036$&  $1.04142\,\pm 0.00058$&  $1.04289\,\pm 0.00036$\\
\hspace{1mm}\\
$\tau$ &   $0.079\,\pm 0.022$& $0.115\,\pm 0.019$&  $0.102\,\pm 0.021$& $0.117\, \pm 0.018$ &   $0.081\,\pm 0.018$& $0.123\,\pm 0.016$&  $0.093\,\pm 0.018$& $0.125\, \pm 0.016$ \\
\hspace{1mm}\\
$n_{\rm s}$ &  $0.968\,\pm 0.012$&$1$  & $0.986\,\pm 0.011$&$1$ &  $0.9666\,\pm 0.0082$&$1$  & $0.9745\,\pm 0.0080$&$1$  \\
\hspace{1mm}\\
$\ln(10^{10}A_{\rm s})$ & $3.094\,\pm 0.043$& $3.166\,\pm 0.036$& $3.137\,\pm 0.042$& $3.168\,\pm 0.034$  & $3.097\,\pm 0.036$& $3.183\,\pm 0.031$& $3.120\,\pm 0.036$& $3.186\,\pm 0.032$ \\
\hspace{1mm}\\
$H_0 /\rm{km \, s^{-1} \, Mpc^{-1}}$ &$67.5\,\pm 1.3$ &$70.35\,\pm 0.59$&$69.6\,\pm 1.1$ &$70.69\,\pm 0.56$ &$67.37\,\pm 0.77$ &$69.92\,\pm 0.45$&$68.30\,\pm 0.73$ &$70.13\,\pm 0.44$ \\
\hspace{1mm}\\
$\sigma_8$ &$0.832\,\pm 0.017$& $0.855\,\pm 0.016$  & $0.841\,\pm 0.018$& $0.853\,\pm 0.015$ &$0.833\,\pm 0.015$& $0.866\,\pm 0.014$  & $0.839\,\pm 0.015$& $0.865\,\pm 0.014$ \\
\hspace{1mm}\\
$Y_{\rm He}$ & $0.251\pm0.021$&$0.296\pm0.010$& $0.273\,\pm 0.019$&$0.294\,^{+0.011}_{-0.009}$ & $0.250\pm0.014$&$0.2929\pm0.0075$& $0.258\,\pm 0.013$&$0.2917\,\pm 0.0073$ \\
\hline
\hspace{1mm}\\
$\bar{\chi}^2_{\rm eff}$ & $11282.87$&$11287.83$&$11289.80$&$11290.05$& $12967.95$&$12982.54$&$12977.10$&$12986.10$ \\
\hline
\end{tabular}}
\end{center}
\caption{68\% credible intervals for $\Lambda$CDM+$Y_{\rm He}$. Planck TT, Planck TT+R16, Planck TTTEEE and Planck TTTEEE+R16 data are considered. The $\bar{\chi}^2_{\rm eff}$ reported are from the corresponding dataset.}
\label{tab:yhe}
\end{table*}

\begin{table*}
\begin{center}\footnotesize
\scalebox{0.74}{\begin{tabular}{c|cccccccc}
\hline
Parameter         & $\Lambda$CDM & $\Lambda$CDM (HZ) & $\Lambda$CDM  & $\Lambda$CDM (HZ) & $\Lambda$CDM & 
$\Lambda$CDM (HZ) & $\Lambda$CDM  & $\Lambda$CDM  (HZ)   \\  
&+$Y_{\rm He}$+$N_{\rm eff}$&+$Y_{\rm He}$+$N_{\rm eff}$&+$Y_{\rm He}$+$N_{\rm eff}$&+$Y_{\rm He}$+$N_{\rm eff}$&+$Y_{\rm He}$+$N_{\rm eff}$&+$Y_{\rm He}$+$N_{\rm eff}$&+$Y_{\rm He}$+$N_{\rm eff}$&+$Y_{\rm He}$+$N_{\rm eff}$\\
\hspace{1mm}\\
       & TT & TT & TT+R16 & TT+R16 & TTTEEE & TTTEEE & TTTEEE+R16 & TTTEEE+R16  \\ 
\hline
\hspace{1mm}\\
$\Omega_{\rm b}h^2$  &  $0.02233\,\pm 0.00037$& $0.02295\,\pm 0.00020$  & $0.02262\,\pm 0.00031$& $0.02295\, \pm 0.00020$ &  $0.02218\,\pm 0.00025$& $0.02300,\pm 0.00014$  & $0.02256\,\pm 0.00022$& $0.02301\, \pm 0.00014$\\
\hspace{1mm}\\
$\Omega_{\rm c}h^2$  &  $0.1201\,^{+0.0069}_{-0.0081}$&  $0.1257\,^{+0.0065}_{-0.0089}$&  $0.1271\,^{+0.0055}_{-0.0062}$&  $0.1246\,^{+0.0049}_{-0.0057}$ &  $0.1156\,^{+0.0044}_{-0.0049}$&  $0.1221\,^{+0.0044}_{-0.0054}$&  $0.1241\,\pm 0.0045$&  $0.1236\,\pm 0.0042$ \\
\hspace{1mm}\\
$\theta_{\rm c}$  &  $1.0411\,\pm 0.0018$&  $1.0404\,^{+0.0020}_{-0.0018}$&  $1.0396\,\pm 0.0015$&  $1.0407\,\pm 0.0012$ &  $1.0420\,\pm 0.0013$&  $1.0413\,\pm 0.0012$&  $1.0401\,\pm 0.0011$&  $1.0409\,\pm 0.0010$\\
\hspace{1mm}\\
$\tau$ &   $0.082\,^{+0.021}_{-0.024}$& $0.109\,\pm 0.018$&  $0.092\,\pm 0.021$& $0.110\, \pm 0.018$ &   $0.078\,\pm 0.018$& $0.118\,\pm 0.016$&  $0.091\,\pm 0.018$& $0.117\, \pm 0.016$ \\
\hspace{1mm}\\
$n_{\rm s}$ &  $0.970\,\pm 0.016$&$1$  & $0.985\,\pm 0.011$&$1$ &  $0.9612\,\pm 0.0096$&$1$  & $0.9780\,\pm 0.0080$&$1$  \\
\hspace{1mm}\\
$\ln(10^{10}A_{\rm s})$ & $3.100\,\pm 0.047$& $3.166\,\pm 0.034$& $3.131\,\pm 0.042$& $3.166\,\pm 0.035$  & $3.085\,\pm 0.038$& $3.181\,\pm 0.031$& $3.126\,^{+0.040}_{-0.035}$& $3.180\,\pm 0.030$ \\
\hspace{1mm}\\
$H_0 /\rm{km \, s^{-1} \, Mpc^{-1}}$ &$67.9\,^{+3.5}_{-4.0}$ &$73.7\,^{+2.2}_{-2.7}$&$72.2\,\pm 1.6$ &$73.4\,\pm 1.4$ &$65.6\,\pm 1.9$ &$70.1\,\pm 1.4$&$68.30\,\pm 0.73$ &$72.3\,\pm 1.1$ \\
\hspace{1mm}\\
$\sigma_8$ &$0.835\,\pm 0.024$& $0.869\,\pm 0.018$  & $0.855\,\pm 0.019$& $0.868\,\pm 0.018$ &$0.822\,\pm 0.018$& $0.873\,\pm 0.015$  & $0.850\,\pm 0.017$& $0.875\,\pm 0.014$ \\
\hspace{1mm}\\
$Y_{\rm He}$ & $0.251\,^{+0.033}_{-0.029}$&$0.252\,^{+0.036}_{-0.030}$& $0.230\,^{+0.031}_{-0.027}$&$0.257\,^{+0.022}_{-0.020}$ & $0.261\,^{+0.019}_{-0.017}$&$0.272\,^{+0.020}_{-0.017}$& $0.242\,\pm 0.019$&$0.265\,\pm 0.015$ \\
\hspace{1mm}\\
$\neff$ & $3.10\,^{+0.50}_{-0.60}$ &$3.75\,^{+0.43}_{-0.57}$& $3.71\,^{+0.31}_{-0.34}$ &$3.67\,^{+0.29}_{-0.33}$& $2.75\,^{+0.29}_{-0.33}$ &$3.42\,^{+0.28}_{-0.33}$ & $3.42\,\pm 0.26$ &$3.54\,\pm 0.25$ \\
\hline
\hspace{1mm}\\
$\bar{\chi}^2_{\rm eff}$ & $11282.87$&$11287.18$&$11286.09$&$11286.90$& $12967.51$&$12982.28$&$12976.41$&$12982.74$ \\
\hline
\end{tabular}}
\end{center}
\caption{68\% credible intervals for $\Lambda$CDM+$Y_{\rm He}$+$N_{\rm eff}$. Planck TT, Planck TT+R16, Planck TTTEEE and Planck TTTEEE+R16 data are considered. The $\bar{\chi}^2_{\rm eff}$ reported are from the corresponding dataset.}
\label{tab:yheneff}
\end{table*}

\subsection{Planck TTTEEE+R16}

The inclusion of R16 data, as in the case of the Planck TT data, has the main effect of favoring a higher Hubble constant and to put an HZ spectrum in better agreement with the data. However, as we can see from Table~\ref{tab:TTTEEE+R16}, now the inclusion of R16 is problematic also for the $\Lambda$CDM+$N_{\rm eff}$ model. Indeed, since the polarization data now better constrains $N_{\rm eff}$ to the standard value, there is now clearly a tension between the datasets even in this scenario. When including R16 we can notice an increase of $\Delta \bar{\chi}^2_{\rm eff}\sim7$ assuming the $\Lambda$CDM+$N_{\rm eff}$ model. As we can see from Table~\ref{tab:yhe} the inclusion of R16 with Planck TTTEEE is even more problematic in case of $\Lambda$CDM+$Y_{\rm He}$ with an increase of $\Delta \bar{\chi}^2_{\rm eff}\sim9$. As we can see, imposing HZ in this case  raises $\bar{\chi}^2_{\rm eff}$ significantly, clearly indicating that HZ no longer provides a good fit. 
Moreover, the assumption of HZ produces a significantly worse fit to the data also in the case of Extended-$10$ and Extended-$11$. We have therefore a higher evidence for HZ with respect to Planck TTTEEE but still worse with respect to the case of the Planck TT+R16 dataset.

\begin{table*}
\begin{center}\footnotesize
\scalebox{0.74}{\begin{tabular}{c|cccccccc}
\hline
Parameter         & $\Lambda$CDM & $\Lambda$CDM (HZ) & $\Lambda$CDM +$N_{\rm eff}$ & $\Lambda$CDM +$N_{\rm eff}$ (HZ) & $\Lambda$CDM & $\Lambda$CDM (HZ) & $\Lambda$CDM +$N_{\rm eff}$ & $\Lambda$CDM +$N_{\rm eff}$ (HZ)   \\  
\hspace{1mm}\\
       & TT+BAO & TT+BAO & TT+BAO & TT+BAO & TTTEEE+BAO & TTTEEE+BAO & TTTEEE+BAO & TTTEEE+BAO  \\ 
\hline
\hspace{1mm}\\
$\Omega_{\rm b}h^2$  &  $0.02227\,\pm 0.00020$& $0.02287\,\pm 0.00020$  & $0.02233\,\pm 0.00024$& $0.02286\, \pm 0.00019$ &  $0.02229\,\pm 0.00014$& $0.02271,\pm 0.00014$  & $0.02229\,\pm 0.00020$& $0.02291\, \pm 0.00013$\\
\hspace{1mm}\\
$\Omega_{\rm c}h^2$  &  $0.1190\,\pm 0.0013$&  $0.11280\,\pm 0.00094$&  $0.1205\,^{+0.0038}_{-0.0042}$&  $0.1300\,\pm 0.0030$ &  $0.1193\,\pm 0.0011$&  $0.11332\,\pm 0.00076$&  $0.1192\,\pm 0.0030$&  $0.1287\,\pm 0.0024$ \\
\hspace{1mm}\\
$\theta_{\rm c}$  &  $1.04096\,\pm 0.00042$&  $1.04175\,\pm 0.00040$&  $1.04081\,\pm 0.00055$&  $1.04000\,\pm 0.00049$ &  $1.04084\,\pm 0.00030$&  $1.04148\,\pm 0.00029$&  $1.04087\,\pm 0.00043$&  $1.03991\,\pm 0.00035$\\
\hspace{1mm}\\
$\tau$ &   $0.081\,\pm 0.018$& $0.140\,\pm 0.017$&  $0.082\,\pm 0.018$& $0.103\, \pm 0.018$ &   $0.082\,\pm 0.016$& $0.141\,^{+0.017}_{-0.015}$&  $0.082\,\pm 0.017$& $0.111\, \pm 0.016$ \\
\hspace{1mm}\\
$n_{\rm s}$ &  $0.9673\,\pm 0.0045$&$1$  & $0.9704\,\pm 0.0088$&$1$ &  $0.9661\,\pm 0.0041$&$1$  & $0.9658\,\pm 0.0076$&$1$  \\
\hspace{1mm}\\
$\ln(10^{10}A_{\rm s})$ & $3.094\,\pm 0.036$& $3.195\,\pm 0.035$& $3.100\,\pm 0.038$& $3.163\,\pm 0.035$  & $3.098\,\pm 0.032$& $3.196\,^{+0.034}_{-0.030}$& $3.098\,\pm 0.035$& $3.176\,\pm 0.031$ \\
\hspace{1mm}\\
$H_0 /\rm{km \, s^{-1} \, Mpc^{-1}}$ &$67.65\,\pm 0.57$ &$70.67\,\pm 0.44$&$68.2\,\pm 1.5$ &$73.04\,\pm 0.62$ &$67.53\,\pm 0.48$ &$70.25\,\pm 0.37$&$67.5\,\pm 1.2$ &$72.63\,\pm 0.54$ \\
\hspace{1mm}\\
$\sigma_8$ &$0.829\,\pm 0.015$& $0.857\,\pm 0.015$  & $0.835\,\pm 0.019$& $0.880\,\pm 0.016$ &$0.832\,\pm 0.013$& $0.860\,\pm 0.014$  & $0.831\,\pm 0.017$& $0.883\,\pm 0.014$ \\
\hspace{1mm}\\
$\neff$ & $3.046$ &$3.046$& $3.14\,\pm 0.23$ &$3.84\,^{+0.13}_{-0.14}$ & $3.046$ &$3.046$& $3.04\,\pm 0.18$ &$3.76\,\pm 0.11$ \\
\hline
\hspace{1mm}\\
$\bar{\chi}^2_{\rm eff}$ & $11286.58$&$11333.60$&$11287.28$&$11297.68$& $12972.07$&$13035.74$&$12973.12$&$12991.85$ \\
\hline
\end{tabular}}
\end{center}
\caption{$68\%$ credible intervals for cosmological parameters for the Planck + BAO.
The $\bar{\chi}^2_{\rm eff}$ reported are from the corresponding Planck+BAO dataset.}
\label{tab:BAO}
\end{table*}

\subsection{Planck + BAO}

We now consider the combination of Planck data with the BAO dataset as used in \cite{planck2015}. 
This dataset is in very good agreement with the Planck $\Lambda$CDM cosmology and we indeed expect a significant exclusion of the HZ spectrum. In Table~\ref{tab:BAO} we report the  $68 \%$ credible intervals for the $\Lambda$CDM and $\Lambda$CDM+$N_{\rm eff}$ models, for the Planck TT+BAO and Planck TTTEEE+BAO datasets.  As we can see the error on $n_{\rm s}$ is further reduced by $\sim 27 \%$ with respect to Planck TT and $\sim 15 \%$ with respect to Planck TTTEEE. The direct consequence for this is that the HZ spectrum worsens the $\bar{\chi}^2_{\rm eff}$ value by $\Delta \bar{\chi}^2_{\rm eff}\sim47$ in case of TT+BAO data
and of $\Delta \bar{\chi}^2_{\rm eff}\sim64$ in case of Planck TTTEEE+BAO.
The situation improves for HZ but not significantly when considering $N_{\rm eff}$. Assuming HZ in $\Lambda$CDM+$N_{\rm eff}$ worsens the $\bar{\chi}^2_{\rm eff}$ value by $\Delta \bar{\chi}^2_{\rm eff}\sim 10$ in case of TT+BAO data
and of $\Delta \bar{\chi}^2_{\rm eff}\sim18$ in case of Planck TTTEEE+BAO.

\begin{table*}
\begin{center}\footnotesize
\scalebox{0.74}{\begin{tabular}{c|cccccccc}
\hline
Parameter         & $\Lambda$CDM & $\Lambda$CDM (HZ) & $\Lambda$CDM +$A_{\rm lens}$ & $\Lambda$CDM +$A_{\rm lens}$ (HZ) & $\Lambda$CDM & $\Lambda$CDM (HZ) & $\Lambda$CDM +$A_{\rm lens}$ & $\Lambda$CDM +$A_{\rm lens}$ (HZ)   \\  
\hspace{1mm}\\
       & TT+WL & TT+WL & TT+WL & TT+WL & TTTEEE+WL & TTTEEE+WL & TTTEEE+WL & TTTEEE+WL  \\ 
\hline
\hspace{1mm}\\
$\Omega_{\rm b}h^2$  &  $0.02233\,\pm 0.00022$& $0.02301\,\pm 0.00020$  & $0.02276\,\pm 0.00028$& $0.02339\, \pm 0.00021$ &  $0.02229\,\pm 0.00015$& $0.02287,\pm 0.00014$  & $0.02245\,\pm 0.00017$& $0.02306\, \pm 0.00015$\\
\hspace{1mm}\\
$\Omega_{\rm c}h^2$  &  $0.1179\,\pm 0.0020$&  $0.1097\,\pm 0.0010$&  $0.1151\,\pm 0.0022$&  $0.1089\,\pm 0.0011$ &  $0.1191\,\pm 0.0014$&  $0.11140\,\pm 0.00086$&  $0.1178\,\pm 0.0015$&  $0.11056\,\pm 0.00085$ \\
\hspace{1mm}\\
$\theta_{\rm c}$  &  $1.04106\,\pm 0.00047$&  $1.04218\,\pm 0.00040$&  $1.04156\,\pm 0.00050$&  $1.04247\,\pm 0.00041$ &  $1.04085\,\pm 0.00032$&  $1.04172\,\pm 0.00029$&  $1.04099\,\pm 0.00033$&  $1.04185\,\pm 0.00029$\\
\hspace{1mm}\\
$\tau$ &   $0.075\,\pm 0.019$& $0.136\,^{+0.018}_{-0.016}$&  $0.055\,^{+0.020}_{-0.022}$& $0.075\, \pm 0.023$ &   $0.074\,\pm 0.017$& $0.139\,\pm 0.016$&  $0.047\,^{+0.018}_{-0.022}$& $0.079\, \pm 0.023$ \\
\hspace{1mm}\\
$n_{\rm s}$ &  $0.9694\,\pm 0.0059$&$1$  & $0.9779\,\pm 0.0066$&$1$ &  $0.9660\,\pm 0.0047$&$1$  & $0.9697\,\pm 0.0047$&$1$  \\
\hspace{1mm}\\
$\ln(10^{10}A_{\rm s})$ & $3.080\,\pm 0.036$& $3.182\,^{+0.036}_{-0.031}$& $3.033\,^{+0.039}_{-0.044}$& $3.059\,\pm 0.045$  & $3.082\,\pm 0.033$& $3.190\,\pm 0.031$& $3.024\,^{+0.037}_{-0.043}$& $3.068\,\pm 0.046$ \\
\hspace{1mm}\\
$H_0 /\rm{km \, s^{-1} \, Mpc^{-1}}$ &$68.11\,\pm 0.92$ &$72.13\,\pm 0.49$&$69.6\,\pm 1.1$ &$72.88\,\pm 0.53$ &$67.58\,\pm 0.64$ &$71.19\,\pm 0.42$&$68.25\,\pm 0.69$ &$71.72\,\pm 0.43$ \\
\hspace{1mm}\\
$\sigma_8$ &$0.820\,\pm 0.014$& $0.838\,\pm 0.014$  & $0.792\,\pm 0.017$& $0.783\,\pm 0.019$ &$0.824\,\pm 0.013$& $0.849\,\pm 0.013$  & $0.797\,^{+0.015}_{-0.017}$& $0.795\,\pm 0.019$ \\
\hspace{1mm}\\
$A_{\rm lens}$ & $1$&$1$& $1.276\,\pm 0.099$&$1.43\,\pm 0.10$ & $1$&$1$& $1.194\,\pm 0.076$&$1.327\,^{+0.081}_{-0.093}$ \\
\hline
\hspace{1mm}\\
$\bar{\chi}^2_{\rm eff}$ & $11312.83$&$11334.25$&$11304.26$&$11312.88$& $12998.66$&$13043.91$&$12992.22$&$13026.54$ \\
\hline
\end{tabular}}
\end{center}
\caption{$68\%$ credible intervals for cosmological parameters for Planck with weak lensing, considering $\Lambda$CDM and $\Lambda$CDM+$A_{\rm lens}$ models.
The $\bar{\chi}^2_{\rm eff}$ reported are from the corresponding Planck+WL dataset.}
\label{tab:WL}
\end{table*}

\subsection{Planck + WL}

As discussed in the introduction the Planck dataset has an internal tension at the level of $2$ standard deviations on the determination of the amplitude of the lensing parameter $A_{\rm lens}$. Interestingly, the inclusion of $A_{\rm lens}$ as a free parameter in the Planck analysis results in a $\sigma_8$ estimate that is in better agreement with the one obtained from cosmic shear surveys. It is therefore important to assess the viability of an HZ model in the framework of a $\Lambda$CDM+$A_{\rm lens}$ model when considering cosmic shear data -  we use the revised version of the CFHTLenS cosmic shear dataset \cite{planck2015}.  
The parameter constraints from Planck TT+WL and Planck TTTEEE+WL data are reported in Table~\ref{tab:WL}. As we can see, comparing with the Planck TT case in Table~\ref{tab:TT} in the standard $\Lambda$CDM case, the inclusion of the WL dataset goes in the direction of slightly increasing $n_{\rm s}$ and lowering $\sigma_8$. HZ spectra are therefore in slightly better agreement with the Planck+WL dataset with respect to the Planck alone data. As we can see, Planck TT+WL suggest an anomalous value for $A_{\rm lens}$ at more than $2.7$ standard deviations. Moreover when the $A_{\rm lens}$ parameter is allowed to vary, $n_{\rm s}$ in the case of Planck TT+WL is now closer to one. When we consider the $6$ parameter HZ model $\Lambda$CDM+$A_{\rm lens}$ we found that this model has in practice the same $\bar{\chi}^2_{\rm eff}$ value of standard $\Lambda$CDM (compare second and fourth column of Table~\ref{tab:WL}). 
The inclusion of the polarization data reduces the uncertainties on $A_{\rm lens}$ but also shifts its value closer to one. For the Planck TTTEEE+WL dataset the indication for $A_{\rm lens}>1$ is now slightly larger than $2.5$ standard deviations. The fact that $A_{\rm lens}$ is now closer to one shifts the value of the spectral index to lower values with respect to the Planck TT+WL case. As a consequence the HZ spectrum is in strong tension with the Planck TTTEEE+WL dataset, increasing $\bar{\chi}^2_{\rm eff}$ by $\sim 28$ even in the $\Lambda$CDM+$A_{\rm lens}$ scenario.

\end{document}